\documentclass[conference]{IEEEtran}

\usepackage{cite}
\usepackage[cmex10]{amsmath}
\usepackage{multirow}

\usepackage{graphicx}
\graphicspath{ {figures/} }

\begin{document}

\title{A Study of Acoustic Features in Arabic Speaker Identification under Noisy Environmental Conditions }

\author{\IEEEauthorblockN{Zhor Benhafid, Kawthar Yasmine Zergat, Abderrahmane Amrouche}
\IEEEauthorblockA{Speech Com \& Signal Proc. Lab.-LCPTS\\Faculty of Electronics and Computer Sciences\\
USTHB, Bab Ezzouar, 16 111, ALGERIA\\
Email: benhafidzhor@hotmail.com, zergatyasmine@gmail.com, abder.amrouche@gmail.com}}

\maketitle

\begin{abstract}
One of the major parts of the voice recognition field is the choice of acoustic features which have to be robust against the variability of the speech signal, mismatched conditions, and noisy environments. Thus, different speech feature extraction techniques have been developed. In this paper, we investigate the robustness of several front-end techniques in Arabic speaker identification. We evaluate five different features in babble, factory and subway conditions at the various signal to noise ratios (SNR). The obtained results showed that two of the auditory feature i.e. gammatone frequency cepstral coefficient (GFCC) and power normalization cepstral coefficients (PNCC), unlike their combination performs substantially better than a conventional speaker features i.e. Mel-frequency cepstral coefficients (MFCC).
\end{abstract}


\IEEEpeerreviewmaketitle

\section{Introduction}
Voice biometry is to identify the speaker using specific information included in speaker speech waves. It can be split into two types: speaker verification (SV) and speaker identification (SI), which can both be text-dependent or text-independent. SV aims to determine if the speaker is who claims to be or not, while speaker identification decides which speaker is talking among a set of speakers.
\par
\begin{figure}[!t]
\centering
\includegraphics [width=3.15in]{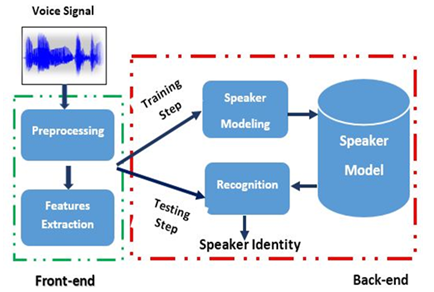}
\caption{ Speaker identification diagram.}
\label{fig:SIS}
\end{figure}

 Speaker identification system (SIS) as every pattern recognition system has two parts shown in Fig.\ref{fig:SIS}: (1) front-end process block where pre-processing and features extraction parameters are done, and (2) back-end process block where pattern classification using speaker modelling, and decision are calculated. However, SIS decreases performance in noisy environments, and mismatched conditions \cite{r1}. Thus, the improvement of robustness of speaker identification system has become a focus question, where much research has been conducted. For example, Missing-data approaches using Gaussian mixture models (GMM) and universal background model (UBM) are used to optimize recognition performance under noisy conditions [\cite{r2}, \cite{r3}. Speech pre-processor such as voice activity detection (VAD) in \cite{r4} has shown robustness against noise. Also, speaker features like gammatone frequency cepstral coefficient, based on an auditory periphery model (CASA) show that this feature captures speaker characteristics and performs substantially better than conventional speaker features in noisy environment \cite{r5}. 
Other studies using GFCC exhibits superior noise robustness to commonly used Mel-frequency cepstral coefficients (e.g. \cite{r6},\cite{r7}, \cite{r8}, \cite{r9}, and \cite{r10}). On the other hand, power normalization cepstral coefficients for speech recognition have shown a substantial improvement in recognition accuracy in the presence of various types of additive noise and reverberant environments \cite{r11} that why paper \cite{r12}, and \cite{r13}proposed a speaker verification system using PNCC combined with other methods which gave good results.
\par This paper investigates the problem of speaker identification using speech samples distorted by additive environmental noises. We review five speech characteristics including MFCC, GFCC, PNCC, perceptual linear predictive (PLP) and line spectral frequency (LSF). The experiments were carried out using the GMM algorithm and Arabic spoken digit (ARADIGIT) database in clean, babble, factory and subway noises conditions. 
\par The rest of the paper is organized as follows: Section II presents the front-end process, the back-end process is given in Section III. Section IV presents the experimental process; results and discussion are given in section V; finally, conclusions are drawn in Section VI.

\section{FRONT-END PROCESS}
\subsection{Mel-frequency cepstral coefficients}
The MFCCs are heuristic representations of acoustic properties, that simulate the human ear. More precisely, they mimic the perceptual representations of acoustic information conveyed by the human auditory system \cite{r14}. The extraction of MFCC features is listed as follow: 
\begin{itemize}
\item Pre-emphasize input speech signal.
\item Calculate short-time Fourier transform (STFT) to get magnitude spectrum.
\item Wrap the magnitude spectrum into Mel-spectrum using tri-angular overlapping windows where centre frequencies of the windows are equally distributed on the Mel scale.
\item Take the log operation on the power spectrum
\item Apply the discrete cosine transform (DCT) on the log-Mel-power- spectrum to derive cepstral features.
The mel-scale calculation is given by the following equation:
\end{itemize}

\begin{equation} \label{eq1}
f_{mel}=2595log_{10}(1+\dfrac{f_{hz}}{700})
\end{equation}

\subsection{Gammatone frequency cepstral coefficients}
GFCCs use technique based on the gammatone filter bank, which attempts to model the human auditory system as a series of overlapping bandpass filters \cite{r8}. GFCCs are calculated as follow:
\begin{itemize}
\item Get gammatone filter bank which impulse response is shown in \ref{eq2}:
\begin{equation} \label{eq2}
g(t)=at^{n-1}e^{-2\pi b_{i} t}cos(2\pi f_{i}t+\varphi)U(t)
\end{equation}
Where  , $f_{i}$ is the center frequency,$ \varphi$ is the phase of carrier, $a$ is the gain of filter, $n$ is the order of filter, $N$ is the number of filter and $b_{i}$ is the attenuation factor that decide the decay rate. The relation between attenuation factor and bandwidth is: $b_{i}=1.019ERE(f_{i})$ where $ERB(f_{i})$ is the equivalent rectangular bandwidth \cite{r10}, it's definition is shown in \ref{eq3}:
\begin{equation}
\label{eq3}
ERB(f_{i}=24.7(\frac{4.37f_{i}}{1000}+1)
\end{equation}
\item Obtain the filter response and decimate it to 100 Hz as a way of time windowing.
\item Apply the cubic root operation on the magnitudes of the decimated outputs. The result provides a T-F decomposition of input which represent a variant of cochleagram.
\item Compute the DCT to have GFCCs. The coefficients are not cepstral coefficients because a cepstral analysis requires a log operation between the first and the second frequency analysis for the deconvolution purpose. Here, we call them cepstral coefficients because of the functional similarities between the above transformation and that of a typical cepstral analysis in the derivation of MFCC \cite{r5}.

\end{itemize}
\subsection{Perceptual linear predictive}
\begin{figure}[!t]
\centering
\includegraphics [width=2.5in]{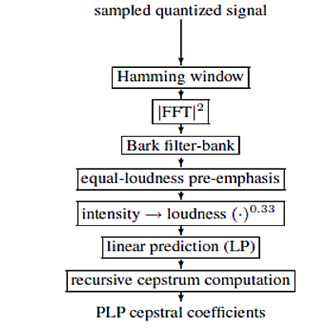}
\caption{ The computation step of PLP \protect\cite{r16}.} 
\label{fig:PLP}
\end{figure}

In PLP technique, several well-known properties of hearing are simulated by practical engineering. The principal of PLP method is illustrated by the abouve block diagram at Fig.\ref{fig:PLP}.

\subsection{Power Normalized Cepstral Coefficients}
The major innovations of PNCC processing include the redesigned nonlinear rate-intensity function, along with
the series of processing elements to suppress the effects of background acoustical activity based on medium-time analysis \cite{r11}. The PNCC process is done in three stages: initial processing, environment processing, and final processing.
\begin{itemize}
\item Apply pre-emphasis followed by STFT on the speech signal and get the short-time spectral power by weighting the magnitude-squared STFT outputs for positive frequencies by the frequency response associated with gammatone filter bank as initial processing.
\item In environmental processing, the main purpose is estimating the noise floor of the speech by temporal asymmetric low pass filter and temporal masking filter, then estimate the “transfer function” by smoothing the ratio of noise floor and mid-time spectral with temporal average, followed by smoothing the short-time spectral using the “transfer function”. These operations called time-frequency normalization processing \cite{r13}.
\item The PNCC vector is obtained in the final process by power-law non-linearity function with exponent 1/15 followed by DCT.
\end{itemize}
\subsection{Line spectral frequency}
\begin{figure}[!t]
\centering
\includegraphics [width=1.5in]{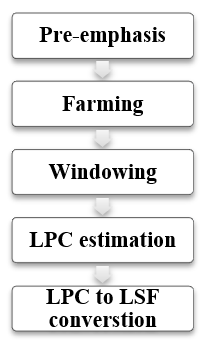}
\caption{ LSF step estimation.}
\label{fig:LSF}
\end{figure}

Line spectral frequencies are an alternative representation to the linear predictive coefficient (LPC) which are the parameters of the vocal tract filter that transformed the excitation signal into a speech signal. The frequency-domain representation of LSFs makes easier the incorporation of human perception system properties. The overall process of the LSF algorithm is shown in the block diagram in Fig.\ref{fig:LSF}.
\section{BACK-END PROCESS}
In this study, we used the state of the art back-end process for speaker recognition i.e. GMM. This feature modelling is a statistical unsupervised machine learning algorithm that uses the statistical variations of the features extracted from the input speech signal to model the speaker. Thus, it provides us with a statistical representation of how a speaker produces sounds; it is a representation of a speaker identity for text-independent speaker recognition\cite{r17}. The d-dimensional Gaussian Probability Density Function (PDF)is given by \ref{eq4}:
\begin{equation} \label{eq4}
N(x|\mu,\Sigma)=\dfrac{1}{(2\pi)^d/2\sqrt|\Sigma|}e^{(-\frac{1}{2}(x-\mu)^{T}\Sigma^{-1}(x-\mu))}
\end{equation} 
Where $\mu$ is the mean and $\Sigma$ is the covariance matrix of the Gaussian. The probability given in a mixture of $K$ Gaussians is:
\begin{equation} \label{eq5}
p(x)=\sum\limits_{j=1}^K w_{j}N(x\mu_{j},\Sigma_{j})
\end{equation}
Where $w_{j}$ is the probability (weight) of the jth Gaussian.  All the weights are positive and sum to one. The Expectation-Maximization (EM) algorithm is used to find GMM parameters i.e. $w_{j}$ , $\mu_{j}$ , $\Sigma_{j}$.
\section{SPEAKER IDENTIFICATION EXPERIMENT}
\subsection{Description of the database}
The database used in this work is an Arabic spoken digit database named ARADIGIT with 62 speakers (31 males and 31 females), recorded in a quiet environment with an ambient noise level below 35 dB, in WAV format, at 22.050 kHz and down-sampled at 16 kHz. It consists of a set of ten digits of the Arabic language (0 to 9), where each speaker repeats the same list 3 times. The speakers are Algerian aged between 18 and 50 years from different regions of Algeria.
\par In our experiments, we used the concatenated sequences of eight numbers (0 to 7) for training and sequences of two numbers (8 to 9) for testing used in \cite{r18}.
\subsection{Expirimental process}
In this study, we reconsidered an SIS in a noisy environment using GMM. First, we modelled speakers at 16 mixtures of GMMs using the training concatenated ARADIGT database. In the testing phase, utterances that are different from the training ones, were digitally mixed at various SNR levels from -6 dB to 18 dB at 6dB intervals using filtering and noise adding tool (FaNT) \cite{r19} and NOISEX’92 database (NATO: AC 243/RSG 10). Three different noise types were used for evaluation: factory noise, speech babble, and subway noise. 
\par In the front-end process, we extracted 39-dimensional auditory features. For each speech frame, we pulled 13 coefficients of PLP, GFCC, and PNCC combined with their first and second derivatives as dynamic features. For the state of the art of speaker, recognition features i.e. MFCC, we took 12 coefficients per frame and its energy which gave a vector of 13 parameters which is concatenated with its delta and Delta delta derivatives. For LSF we extracted the commonly used number of parameters i.e. 10.
\par As the second step we evaluated the combination of the more robustness parameter. For that, we combined 26-dimensional features of GFCC with PNCC, PLP with GFCC, and PLP with PNCC.
\par For evaluating performances of the proposed Arabic speaker identification system, we used the identification rate accuracy (IR) given by:
\begin{equation} \label{eq6}
IR=\dfrac{number of correct identifications}{total number of trails}*100
\end{equation}
\section{RESULTS AND DISCUSSION}

\begin{table*}[!t]
\renewcommand{\arraystretch}{1.7}
\caption{speaker identification accuracy ($\%$) in three environmental noisy conditions}
\label{table1}
\centering
\begin{tabular}{| c | c | c | c | c | c | c |}
\hline
 \multirow{2}{*}{\bfseries Noise} & \multirow{2}{*}{ \bfseries SNR} & \multicolumn{5}{|c|}{Acoustic Features } \\ \cline{3-7} 

 &  &  MFCC &  LSF & PLP & PNCC &  GFCC  \\
\hline\hline
\multirow{5}{*}{\textbf{Babble}}
& -6 & 4,83 & 9,67 & 6,45 & 6,45  & 12,90 \\
& 0 & 12,90 & 19,35 & 16,12 & 24,19 & 24,19 \\
& 6 & 20,96 & 45,16 & 45,16& \textbf{70,96}&\textbf{72,58}   \\
& 12 & 38,70 & \textbf{64,51} & \textbf{91,93} &\textbf{91,93}  & \textbf{95,16}\\
& 18& \textbf{72,58}&\textbf{77,41}  &\textbf{98,38}  & \textbf{95,16} & \textbf{100} \\
\hline\hline
\multirow{5}{*}{\textbf{Factory}}
& -6 & 1,61 & 1,61 & 1,61 & 4,83 & 4,83\\
& 0 & 4,83 & 4,83 & 4,83 & 11,29 & 20,96 \\
& 6 & 9,67 & 19,35 & 16,12 & 48,38  & \textbf{62,9032} \\
& 12 & 19,35 & 30,64 & 45,16 & \textbf{75,80} &\textbf{93,54} \\
& 18 & \textbf{ 53,22}&  \textbf{53,22}& \textbf{87,09} & \textbf{88,70}  & \textbf{100}\\
\hline\hline
\multirow{5}{*}{\textbf{Subway}}
& -6 & 4,83 & 1,61  & 1,61  & 4,83 & 1,61 \\
& 0 & 6,45 & 1,61 & 1,61 & 6,45 & 3,22 \\
& 6 & 16,13 & 1,61 & 1,61 & 29,03 & 17,74 \\
& 12 & 27,41 & 1,61 & 19,35 &  \textbf{53,22}& \textbf{62,90} \\
& 18 & \textbf{ 53,22}& 8,06  & 40,32 &\textbf{77,41}  & \textbf{88,70} \\

\hline
\end{tabular}
\end{table*}
Table \ref{table1} shows the Arabic SI performances in babble, factory and subway noisy environments conditions using MFCC, GFCC, PLP, and PNCC as auditory features and LSF. From the obtained results, it can be said that the GFCCs combined with their dynamic information give the best performances in the three different noisy environments, followed by PNCCs in the second place. It depends mainly on the use of gammatone filter bank for both GFCC and PNCC. Also, the implementation details such as the use of the cubic root of the decimated filter response in the GFCC process and estimating the noise by medium temporary analysis then reducing the effect of noise by the short time spectral in PNCC computation, improve the SI performances. The experiments show also that:
\begin{itemize}
\item Each noisy environment influences the robustness of speaker identification differently; subway noise is the one that most decreases the speaker identification rate accuracy.
\item LSF and MFCC have less performance in the three environmental conditions. However, MFCC is better than LSF in subway noise.
\item PLPs start to give good performances at 12 dB in babble and factory noise, contrary to MFCC and LSF. This is because PLP like GECC take into account that the perceived loudness is a cubic root of voice intensity. 
\item It is noticed that the Arabic speaker identification using the GMM gives the best performances using the GFCC followed by PNCC and PLP at third position. However, their combination doesn’t provide more robustness for identification as it is shown in Fig.\ref{fig:results}.
\begin{figure*}[!th]
\centering
\includegraphics [width=6.5in]{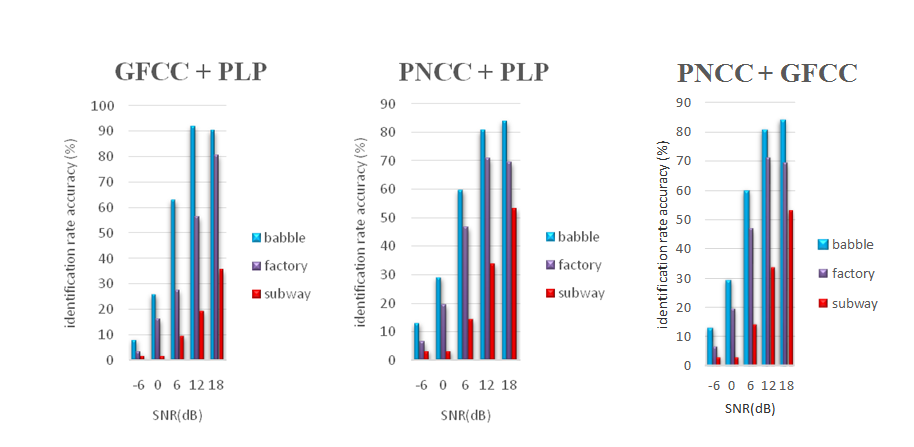}
\caption{SID accuracy ($\%$) comparisons of the proposed combined system.}
\label{fig:results}
\end{figure*}
\end{itemize}

\section{Conclusion}

This work describes the evaluation of speaker identification system using different speech features in babble, factory, and subway noisy environmental conditions. We investigated the MFCCs, PLPs, GFCCs and PNCCs combined with their dynamic characteristics as auditory features and LSF with an Arabic database using GMM.
The obtained results showed that GFCC and PNCC are the most robust features against environmental noise, unlike their combination.
As future work, we recommend evaluating the robustness of the GFCC and PNCC in environmental noisy conditions to discriminate Arabic speakers using a deep neural network classifier in the verification tasks.
\bibliographystyle{IEEEtran}
\bibliography{references}

\end{document}